\documentclass{article}
\usepackage[final]{neurips_2020}
\usepackage{microtype}
\usepackage{graphicx}
\usepackage{booktabs} 
\usepackage{natbib}
\usepackage{hyperref}
\newcommand*\diff{\mathop{}\!\mathrm{d}}
\usepackage{amsmath}
\usepackage{amsfonts}
\usepackage{subcaption}
\usepackage{siunitx}

\usepackage[utf8]{inputenc} 
\usepackage[T1]{fontenc}    
\usepackage{hyperref}       
\usepackage{url}            
\usepackage{booktabs}       
\usepackage{amsfonts}       
\usepackage{nicefrac}       
\usepackage{microtype}      

\title{Gene Regulatory Network Inference with Latent Force Models}

\author{%
  Jacob Moss \\
  University of Cambridge,\\
    Computer Laboratory,\\
    Cambridge CB3 0FD, UK\\
  \texttt{jm2311@cam.ac.uk} \\
   \And
   Pietro Li\'{o} \\
    University of Cambridge,\\
    Computer Laboratory,\\
    Cambridge CB3 0FD, UK \\
   \texttt{pl219@cam.ac.uk} \\
}

\begin{document}

\maketitle

\begin{abstract}
Delays in protein synthesis cause a confounding effect when constructing Gene Regulatory Networks (GRNs) from RNA-sequencing time-series data. Accurate GRNs can be very insightful when modelling development, disease pathways, and drug side-effects. We present a model which incorporates translation delays by combining mechanistic equations and Bayesian approaches to fit to experimental data. This enables greater biological interpretability, and the use of Gaussian processes enables non-linear expressivity through kernels as well as naturally accounting for biological variation. 
\end{abstract}

\section{Introduction}
\label{submission}
Scarcity of data, whether due to unobservable components, low-frequency sampling, or inconsistencies (such as varying results across cell lines), poses a major challenge to modelling biological systems. In this work we therefore look at introducing stronger model structures in order to cope in a low-data regime. The approach combines data-driven and mechanistic equations yielding biologically interpretable inferences of transcriptional regulation over time and translational delays. Such models can rely on a mechanistic component in low-data or uncertain situations.

When modelling transcriptional regulation, a sensible assumption is that the production rate of gene $j$'s mRNA, $m_j(t)$, is in some way dependent on the abundances and types of transcription factors (TFs) present in the cell, $f_i(t)$. Post-transcriptional regulation also occurs, such as RNA splicing, translation, and protein processing events. Delays are rampant; translation occurs at different rates for different mRNAs, and ribosomes can pause synthesis. In this paper we ask whether incorporating delays in the mechanism can better explain transcriptomic data. We do this by treating the individual TF protein activities, $p_i(t)$, as a delayed response to transcription. 

\begin{figure}[t]
    \centering
    \includegraphics[width=9.5cm]{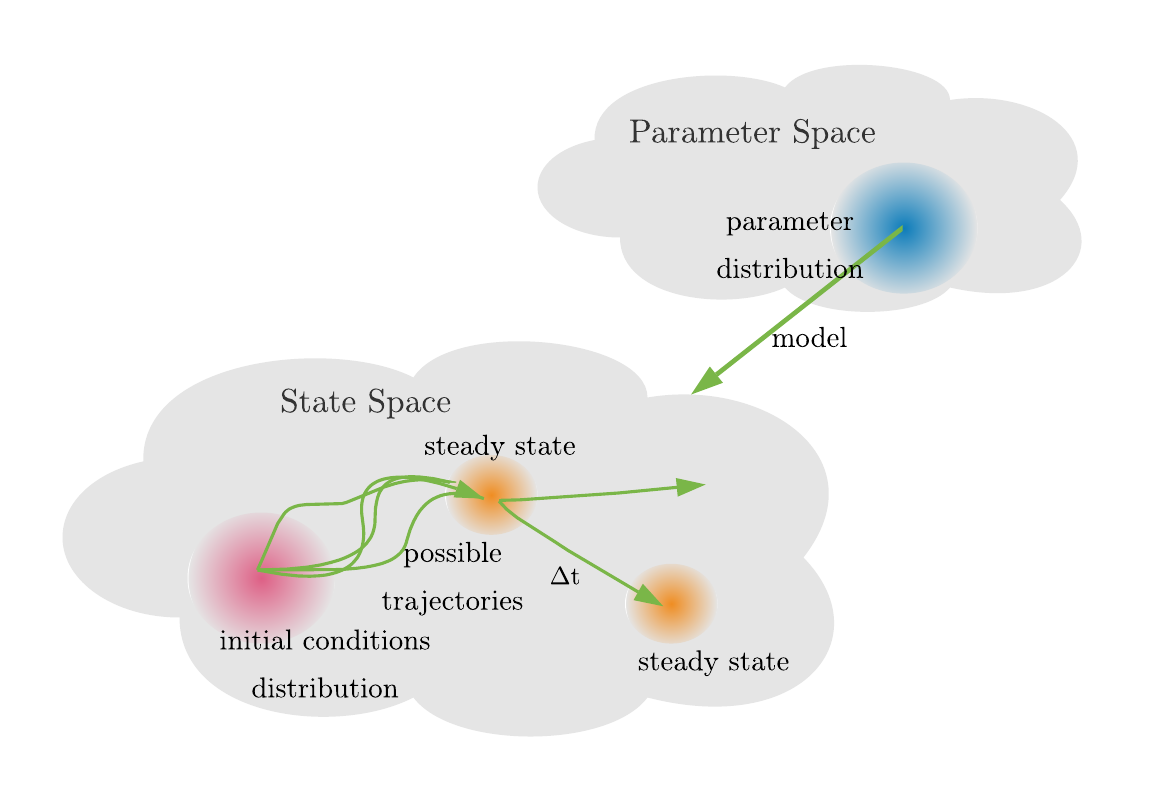}
    \caption{Bayesian dynamical system. A distribution of parameters determines the dynamics of the system. Trajectories lead toward unstable steady states with possible delays and periodicity. Adapted from \cite{gunawardena2010models}.}
    \label{fig:y equals x}
    \vskip -0.1in
\end{figure}

The GRN inference problem can be posed as a series of regressions where the weights represent the influence each TF has on the gene expression level. Selecting from the infinite possible basis functions is challenging, and current solutions tend to restrict the choices to some \textit{a priori} selection (e.g. see \cite{singh}), which requires unreasonable domain knowledge.

Typically, researchers have assumed that the basis functions are identical and linear and often use a metric like Pearson's correlation. This is justified because genes that are correlated are likely to be co-expressed. However, due to effects such as pleiotropy, the assumption of identical and linear basis functions is questionable. In fact, \citet{dutil} showed experimentally that there is non-linearity in gene expression data. This paper tackles the intractable problem without requiring such substantial assumptions.

\subsection{Time Delays in Gene Regulation}\label{sec:timedelay}
Gene expression is an inherently dynamic process, and RNA-seq data only gives us a snapshot of the system in time. It has been shown that there are often long delays after transcription before mRNA is produced \citep{hao2013rna}. Modelling these delays is crucial for determining event \textit{causality}, e.g. the driver of a particular translation. 

ChIP-seq and RNA-seq data can be combined to find protein binding sites and thereby determine how proteins regulate gene expression. The data has been used to show that 11\% of genes exhibit transcriptional delays of more than 20 minutes, and these genes tend to be short \cite{honkela2015genome}.

There are also delays in the translation phase. Intuitively there should be a correlation between the mRNA level and its associated protein, but due to delayed protein synthesis this is not always the case \cite{gedeon2012delayed}. There are several causes of such delays, such as ribosomal pausing and delayed folding. This paper tackles this problem by incorporating a delay parameter in a translation ordinary differential equation (ODE). 

\subsection{Latent Force Models}\label{sec:diffeq}
Latent Force Models (LFMs) \citep{alvarez2009latent} involve a combination of mechanistic and data-driven approaches. These models have been used with success for modelling transcriptional regulation \citep{lawrence2007modelling,titsias2012identifying,honkela2015genome} and post-transcriptional regulation \citep{lopez2019physically}.

\citet{barenco2006ranked} proposed a novel approach involving Markov chain Monte Carlo (MCMC) which they applied to microarray time-series data on the p53 transcription factor network, acquired by irradiating a human leukemia cell line. The mRNA expression levels were modelled by an ODE in terms of the basal transcription rate, decay rate, and sensitivity to transcription. The state of the art prior to this study had been methods based on correlation.

Following from this work, \citet{lawrence2007modelling} used the linearity of covariance to show that applying a Gaussian process prior to the latent function $f(t)$ yields a covariance function over the gene mRNA levels that can be computed analytically. With a linear response this results in a tractable predictive posterior since the integrand is a linear operator. The kernel choice determines shape of time series. 

\section{Non-Linear Response \& Multiple TFs}
Only a few families of differential equations and kernels result in analytically solvable solutions. The aforementioned models are limited to the single TF and linear responses. Although the MAP-Laplace approximation can be used to get around this \cite{lawrence2007modelling}, we follow instead a similar formulation to \citet{titsias2012identifying}, who developed an MCMC algorithm incorporating non-linear responses, multiple transcription factors, and a translation mechanism. 

\subsection{Mechanistic Component}
Each gene's transcriptional regulation is modelled by
\begin{equation*}
\vspace{-3mm}
    \frac{dm_j(t)}{dt}=b_j+s_jG(p_1(t),...,p_I(t);\textbf{w}_j,w_{j0}) - d_jm_j(t),
\end{equation*}

with initial conditions $a_j$, basal transcription rate $b_j$, decay rate $d_j$, sensitivity $s_j$, non-linear response $G(\cdot)$ allowing transcription factors to cooperate or compete when affecting the transcription, and $\mathbf{w_j}= [w_{j,i}]_{i=1}^I$ are the weights denoting the strength of interaction between the $j$-th gene and the $i$-th transcription factor. These weights are required to construct the GRN. We can solve the ODE, yielding
\begin{equation*}
m_j(t) = \frac{b_j}{d_j} + \left(a_j-\frac{b_j}{d_j} \right)e^{-d_jt} + s_j\int^t_0 G(\mathbf{p}(t);\textbf{w}_j)e^{-d_j(t-u)} du,
\end{equation*}
The 1-D convolution integral here, however, is not tractable, so we use a numerical approximation like the trapezoidal rule. We have data for the TF mRNA concentration, $f_i(t)$, so we express $p_i(t)$, the TF protein activity, using another ODE model,
\[\frac{\diff p_i(t)}{\diff t} = f_i(t-\Delta_i)-\delta_ip_i(t),\]

where $\delta_i$ is the protein decay rate and $\Delta_i$ is a translation delay parameter. This has the solution
\[p_i(t) = \int^t_0f_i(u-\Delta)e^{-\delta_i(t-u)}\diff u + a_ie^{-\delta_i t},\]
where $a_i$ are the initial conditions. All that is left to define for the mechanistic component is the 
the form of $G(\cdot)$, the response to transcription factors. To capture the saturating aspect of transcription factor proteins a sigmoid function can be justified. Indeed, GRNs are commonly modelled as switch-like networks (see \citet{giampieri2011stochastic}), which suggests a sigmoidal function would be appropriate. The weights capture levels of collaboration or competition between the TFs. Thus we give $G(\cdot)$ the form
\[G(\mathbf{p}(t);\textbf{w}_j)=\frac{1}{1+\exp\left(-w_{j0}-\sum^I_{i=0}w_{ji}\log p_i(t)\right)},\]
where $w_{j0}$ is a bias. We can see how it emerges from the Hill function with Hill coefficients $\mathbf{w}_j$:
\begin{equation*}
    \begin{split}
        G(\mathbf{p}(t);\textbf{w}_j) &= \frac{\prod_i p_i(t)^{w_{ji}}}{\prod_i p_i(t)^{w_{ji}} + e^{-w_{j0}}}\\&=\frac{\exp(\sum w_{ji}\log p_i(t))}{\exp\left(\sum w_{ji}\log p_i(t)\right)+\exp(-w_{j0})}
    \end{split}
\end{equation*}
\citet{santillan2008use} presents a more detailed discussion of the use of Hill kinetics for modelling transcriptional regulation.

\subsection{Probabilistic Component}
Precursor models such as the one defined by \citet{titsias2012identifying} used the Metropolis-Hastings algorithm for each hyperparameter and latent function. Our model instead uses a mixture of sampling algorithms, most of which use automatic differentiation from Tensorflow to take gradient-informed steps, thereby enabling gradient-based samplers, such as Hamiltonian Monte Carlo. 

We use the No U-Turn Sampler (NUTS) \citep{hoffman2014no} to sample kinetic parameters $\mathbf{k}^m_j=(a_j,b_j,s_j,d_j)$, $\mathbf{k}^f_i=(a_i,\delta_i)$, $\mathbf{w}_j$, $w_{j0}$, $l_i^2$, and $v_i$. NUTS is a variant of the Hamiltonian Monte Carlo algorithm which adaptively chooses the number of steps to take, thereby avoiding the need for tuning. $\Delta_i$ is a discrete random variable since we have discretised time. Given the relatively small sample space we can compute the complete normalised posterior, lending itself to inverse transform sampling which uses the inverse cumulative distribution function. $f_i(t)$ is sampled using a Metropolis-Hastings sampler similar to that described in \citet{titsias2012identifying}. The noise variances, $\sigma^2_m,\sigma^2_f$ were assigned conjugate priors which enabled Gibbs sampling. 

\subsubsection{Likelihood}
We assign a Gaussian likelihood to the data\[\log p(\mathbf{\widetilde  M|M},\boldsymbol{\theta}) = \sum_j^J\sum^{N_m}_t \log\mathcal{N}(\widetilde m_j(t)|m_j(t), \sigma^2_j(t)),\]
where $\boldsymbol{\theta}$ represents all the mechanistic parameters for $J$ genes with $N_m$ observations each. Before we can calculate $m_j(t)$ and $f_i(t)$, we must specify \textit{a priori} the granularity with which to predict our latent functions, i.e. the number of points to predict in between the range of observation times. Since we compute the likelihoods at the observation timepoints, the discretised time grid must contain these points. An easy way to do this is by generating a linearly spaced vector of length $f(n, d)$ where \(f(n, d) = (n-1)(d-1)+1,\) where $n$ is the number of augmented timepoints and $d$ is the discretisation multiple, which we set to $10$ to obtain $10\times$ the granularity of the observations. We denote this time vector $\boldsymbol{\tau}$.

As mentioned, we use the trapezoidal rule for approximating the convolution integrals, e.g. for the $\mathbf{p}_i$ vector
\begin{equation*}
    \int^t_0f_i(t)e^{-\delta_i(t-u)}\diff u \approx \Delta \tau \frac12\sum_{k \in \boldsymbol{\tau}}\left(f_i( \boldsymbol{\tau}_{k-1})e^{-\delta_i(t-\boldsymbol{\tau}_{k-1})} + f_i(\boldsymbol{\tau}_k)e^{-\delta_i(t- \boldsymbol{\tau}_k)}\right),
\end{equation*}
where $\Delta \tau=\boldsymbol{\tau}[1]-\boldsymbol{\tau}[0]$. The same is then done for the gene ODE, yielding $m_j(k)$ for all $k \in \boldsymbol{\tau}$.

\subsubsection{Prior Choices}\label{sec:prior}

\begin{table}[t]
\caption{Prior distributions for parameters. $Logit$-$\mathcal{N}(a, b)$ denotes a zero mean, 1.8 variance logit-normal distribution bounded by $a$ and $b$. $\Delta_{\text{min}} = \min(t_{i}-t_{i-1}|0<i<N)$ }
\label{tab:modelcompare}
\vskip 0.15in
\begin{center}
\begin{small}
\begin{sc}
\begin{tabular}{lp{40mm}}
\toprule
Parameter & Prior \\
\midrule
$f_i(t)$           & $\mathcal{N}(\mathbf{0}, \kappa_i(t,t'))$ \\
$\mathbf{k}^m_j$         & $Logit$-$\mathcal{N}(0.05, 8)$ \\
$\mathbf{k}^f_i$         & $Logit$-$\mathcal{N}(0.05, 8)$ \\
$\mathbf{w}_j$, $w_{j0}$ & $Logit$-$\mathcal{N}(-1, 1)$  \\
$\sigma^2_m, \sigma^2_f$             & $InvGamma(0.01, 0.01)$  \\
$l_i^2$                  & $Logit$-$\mathcal{N}(\Delta_{\text{min}}, t_N)$  \\
$v_i$                    & $Logit$-$\mathcal{N}(0.01, 0.01)$  \\
$\Delta$                 & $Discrete$-$\mathcal{U}(0, 10)$ \\
\bottomrule
\end{tabular}
\end{sc}
\end{small}
\end{center}
\vskip -0.1in
\end{table}

Tuning the step sizes of such a complex model would be computationally intensive, so we choose a global isotropic step size. One method of doing this is by applying a sigmoid bijector to get all parameters in the range $(0,1)$. We can impose a Gaussian prior on the pre-transformed variables by way of the $Logit$-$\mathcal{N}(a, b)$ (logit-normal) prior. We used this technique for all parameters except the latent functions, noise variances, and $\Delta$. The step sizes were chosen to achieve at least 80\% acceptance rate for the group sampled with NUTS, and 25\% for the latent functions.

We parameterise the logit-normal with a mean and variance of $0$ and $1.8$ respectively resulting in a roughly uniform bounded distribution between $a=0$ and $b=1$. These bounds are expanded to arbitrary $a$ and $b$ by applying $x := \frac{x-a}{b-a}$ before passing into the density function. 

We place a positivity constraint for our latent function (negative gene expression levels are nonsensical) by using the transformation $\log(\exp(f_i(t))-1))$. The latent forces $f_i(t)$ are assigned a zero mean Gaussian process (GP) prior with a radial basis function (RBF) covariance function

\[\kappa_i(t,t') = v_i\exp\left(-\frac{1}{2l^2_i}(t-t')^2\right),\]
where $v_i$ is a scale factor indicating the average deviation from the mean and $l$ is the lengthscale determining the waviness of the function. If problematic, the oscillatory nature of the RBF kernel can be countered by using the multilayer perceptron (MLP) kernel \citep{lawrence2007modelling}. 

The priors on the decay rate parameters correspond to an mRNA half-life with quartiles:
\begin{equation*}
    t_{\frac12}^{5\%} = \frac{\ln 2}{8} = 5.2 \text{min}, \quad
    t_{\frac12}^{95\%} = \frac{\ln 2}{0.05} = 13.8 \text{h}
\end{equation*}

The interaction weights, $\mathbf{w}_j$, were given a range $[-1, 1]$ since they can take positive and negative values indicating activation and repression respectively. The gene and TF mRNA observation for the variances are assigned conjugate inverse Gamma priors. These are assigned vague concentration and scale hyperparameters.

\section{Results}
\begin{figure}[t]
\vskip 0.2in
     \centering
     \begin{subfigure}[b]{0.49\linewidth}
         \centering
         \includegraphics[width=6cm]{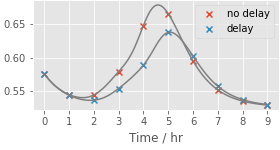}
         \caption{One of the 5 artificial genes.}
         \label{fig:y equals x}
     \end{subfigure}
     \hfill
     \begin{subfigure}[b]{0.49\linewidth}
         \centering
         \includegraphics[width=6cm]{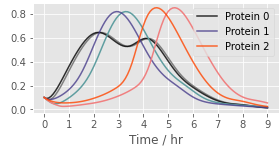}
         \caption{3 artificial TF proteins.}
         \label{fig:five over x}
     \end{subfigure}
        \caption{Examples from the artificial dataset.}
        \label{fig:artif}
        \vskip -0.2in
\end{figure}

In order to demonstrate the ability of the model to predict translational delays we created an artificial dataset of 5 genes and 3 TFs, with 10 observations over the course of 8 hours. We imposed different delays on the transcription factor proteins; TF 0 had a delay of 6 min, TF 1 had 24 min, and TF 2 had 1 hr. The effects of such a delay on the genes is shown in Figure \ref{fig:artif}.

\begin{figure}[b]
\vskip 0.2in
     \centering
     \begin{subfigure}[b]{0.49\columnwidth}
         \centering
         \includegraphics[width=5cm]{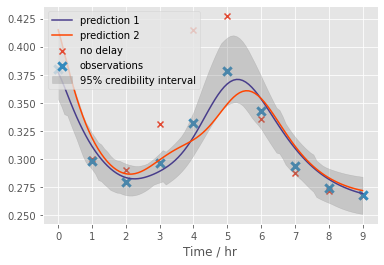}
         \caption{Two predictions with different delay modes.}
         \label{fig:modes}
     \end{subfigure}
     \hfill
     \begin{subfigure}[b]{0.49\columnwidth}
         \centering
         \includegraphics[width=5cm]{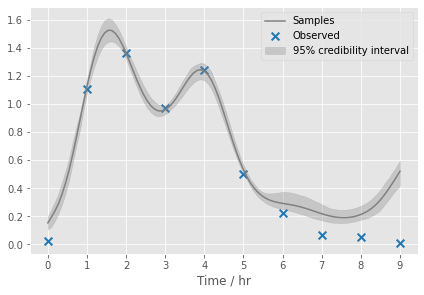}
         \caption{One of the inferred transcription factor mRNAs.}
         \label{fig:infertrans}
     \end{subfigure}

        \caption{Inferred gene and transcription factor.}
        \label{fig:infer}
        \vskip -0.2in
\end{figure}

\begin{figure}[t]
\vskip 0.2in
     \centering
     \begin{subfigure}[b]{0.245\linewidth}
         \centering
         \includegraphics[width=\textwidth]{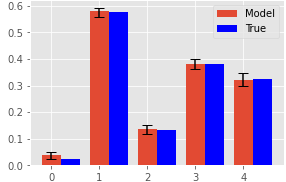}
         \caption{Initial conditions.}
         \label{fig:y equals x}
     \end{subfigure}
     \hfill
     \begin{subfigure}[b]{0.245\linewidth}
         \centering
         \includegraphics[width=\textwidth]{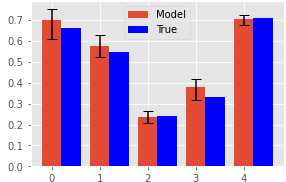}
         \caption{Basal rates.}
         \label{fig:five over x}
     \end{subfigure}
      \begin{subfigure}[b]{0.245\linewidth}
         \centering
         \includegraphics[width=\textwidth]{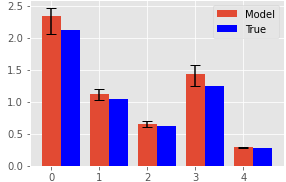}
         \caption{Decay rates.}
         \label{fig:y equals x}
     \end{subfigure}
     \hfill
     \begin{subfigure}[b]{0.245\linewidth}
         \centering
         \includegraphics[width=\textwidth]{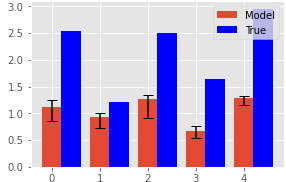}
         \caption{Sensitivities.}
         \label{fig:five over x}
     \end{subfigure}

    \caption{Inferred transcription parameters compared to the artificial parameters. Gene numbers are along the x-axis.}
    \label{fig:genek}
        \vskip -0.2in
\end{figure}

\begin{figure}[t]
\vskip 0.2in
     \centering
     \begin{subfigure}[b]{0.47\linewidth}
         \centering
         \includegraphics[width=4cm]{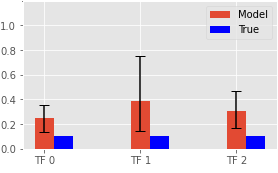}
         \caption{Initial conditions.}
         \label{fig:y equals x}
     \end{subfigure}
     \hfill
     \begin{subfigure}[b]{0.47\linewidth}
         \centering
         \includegraphics[width=4cm]{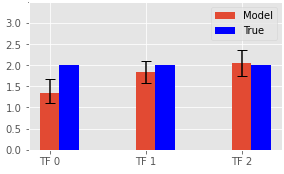}
         \caption{Sensitivities.}
         \label{ss}
     \end{subfigure}

        \caption{Inferred translation parameters compared to the artificial parameters.}
        \label{fig:tfk}
        \vskip -0.2in
\end{figure}

A single chain was sampled for 1,500 iterations with a burn-in of $200$. We chose a granularity of 100 points and global NUTS step size of $\num{1e-5}$. We plotted the converged parameter values with 95\% credible intervals in Figures \ref{fig:genek} and \ref{fig:tfk}. The decay parameters demonstrated unstable convergence behaviours suggesting the presence of multiple optima. This is likely given that we are using a small dataset of only 5 genes. In order to test this theory we plotted two such optima in Figure \ref{fig:modes}. Both give good fits for the delayed observations (blue points), but use different delays.

The sampler converges for all latent functions to a good empirical fit. Figure \ref{fig:infertrans} shows that the zero mean of the GP prior under the positivity transformation causes an upward tail in transformed space. A mean function could alleviate this issue.

\begin{figure}[b]
     \centering
     \begin{subfigure}[b]{0.49\linewidth}
         \centering
         \includegraphics[width=6cm]{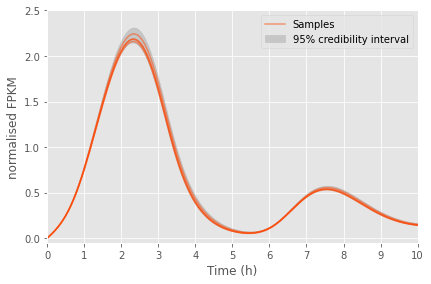}
         \caption{Expression profile of p53 protein inferred by our model.}
         \label{fig:infer_blottings}
     \end{subfigure}
     \hfill
     \begin{subfigure}[b]{0.49\linewidth}
         \centering
         \includegraphics[width=6cm]{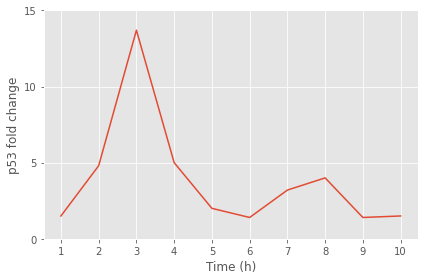}
         \caption{Fold change from the western blot in \citet{hafner2017p53}.}
         \label{fig:blottings}
     \end{subfigure}

        \caption{Inferred p53 profile alongside the results of an abundance quantification experiment.}
        \label{fig:p53}
        \vskip -0.2in
\end{figure}

We used the RNA-seq dataset of p53-regulated genes from \citet{hafner2017p53} to demonstrate the models inferential ability. In this time-series dataset, the MCF7 cell line is irradiated with 10 Gy and samples are taken every hour for 12 hours. In order to measure the abundance of p53 protein, \citeauthor{hafner2017p53} used western blotting, a wet-lab test. The p53 regulatory network is important to understand since p53 acts as a tumour suppressor: it is activated in response to DNA damage, such as in cancer. After a radiation dose there is typically a series of characteristic p53 activation pulses. This is shown in the western blot result in Figure \ref{fig:blottings}. 

The inferred p53 protein is plotted in Figure \ref{fig:infer_blottings} alongside the quantification of the western blot, clearly illustrating the model's ability to infer a latent force through an ODE cascade. It is important to note that the model did not have access to the blot quantification or p53 concentration, it only had access to a group of 22 genes (corresponding to cluster 1 in \citet{hafner2017p53}).

We found it necessary to use an exponential transform of the kinetic parameters to account for numerical issues in the logistic-normal distribution. This was due to the large range of peak heights after data normalisation. The transform enabled the model to search a much larger parameter space.

\begin{figure}[t]
    \centering
    \includegraphics[width=8cm]{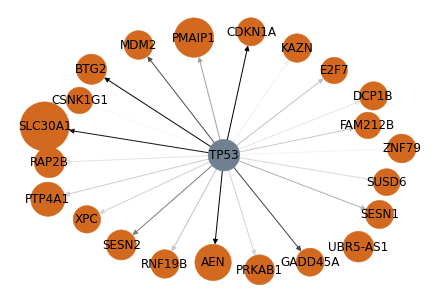}
    \caption[Gene Regulatory Network for the p53 TF.]{Gene Regulatory Network for the p53 TF (centred), where the size of each node indicates the basal transcription rate, and the opacity of each edge is the sensitivity to transcription.}
    \label{fig:hp53-grn}
\end{figure}

Since this is a single input module motif (where we have one TF), the weights are disabled for identifiability purposes. In this case, a GRN can be constructed using the sensitivity parameters representing the strength of regulation. This is shown in Figure \ref{fig:hp53-grn}, from which one can conclude that the most sensitive targets of p53 are CDKN1A, AEN, SLC30A1, BTG2, GADD45A, and MDM2. 

\section{Evaluation}
The model is able to find good parameter estimates for the artificial dataset. On real datasets, the model exhibits excellent inferential ability. Furthermore, the GRN can be verified from the literature. Many of the targets have known associations with p53: CDKN1A codes for p21, key to preventing cell division; AEN induces apoptosis (cell death) and is regulated by p53 phosphorylation; and BTG2 is a key component of the cellular response to DNA damage. The strong sensitivity of SLC30A1 may be worth further investigation; genetic mutations may cause hypermanganesemia, which is associated with neurological disorders like Parkinson's. Finally, MDM2 is a post-translational regulator of p53; a process that could be modeled in future work.

As mentioned in Section 1, there are also transcriptional delays; the problem of over-parameterisation would need to be addressed if decay variables are also used in the transcription ODE. There are also multiple possible solutions for the delay parameters. Future work should look at including more gene mRNA data to yield a clearer optimum, and using multiple chains to improve convergence diagnostics.

\newpage


\bibliography{example_paper}

\begin{thebibliography}{15}
\providecommand{\natexlab}[1]{#1}
\providecommand{\url}[1]{\texttt{#1}}
\expandafter\ifx\csname urlstyle\endcsname\relax
  \providecommand{\doi}[1]{doi: #1}\else
  \providecommand{\doi}{doi: \begingroup \urlstyle{rm}\Url}\fi

\bibitem[Gunawardena(2010)]{gunawardena2010models}
Jeremy Gunawardena.
\newblock Models in systems biology: the parameter problem and the meanings of
  robustness.
\newblock \emph{Elements of Computational Systems Biology. New Jersey: John
  Wiley and Sons}, pages 21--48, 2010.

\bibitem[Singh and Vidyasagar(2015)]{singh}
Nitin Singh and Mathukumalli Vidyasagar.
\newblock blars: an algorithm to infer gene regulatory networks.
\newblock \emph{IEEE/ACM transactions on computational biology and
  bioinformatics}, 13\penalty0 (2):\penalty0 301--314, 2015.

\bibitem[Dutil et~al.(2018)Dutil, Cohen, Weiss, Derevyanko, and Bengio]{dutil}
Francis Dutil, Joseph~Paul Cohen, Martin Weiss, Georgy Derevyanko, and Yoshua
  Bengio.
\newblock Towards gene expression convolutions using gene interaction graphs.
\newblock \emph{arXiv preprint arXiv:1806.06975}, 2018.

\bibitem[Hao and Baltimore(2013)]{hao2013rna}
Shengli Hao and David Baltimore.
\newblock Rna splicing regulates the temporal order of tnf-induced gene
  expression.
\newblock \emph{Proceedings of the National Academy of Sciences}, 110\penalty0
  (29):\penalty0 11934--11939, 2013.

\bibitem[Honkela et~al.(2015)Honkela, Peltonen, Topa, Charapitsa, Matarese,
  Grote, Stunnenberg, Reid, Lawrence, and Rattray]{honkela2015genome}
Antti Honkela, Jaakko Peltonen, Hande Topa, Iryna Charapitsa, Filomena
  Matarese, Korbinian Grote, Hendrik~G Stunnenberg, George Reid, Neil~D
  Lawrence, and Magnus Rattray.
\newblock Genome-wide modeling of transcription kinetics reveals patterns of
  rna production delays.
\newblock \emph{Proceedings of the National Academy of Sciences}, 112\penalty0
  (42):\penalty0 13115--13120, 2015.

\bibitem[Gedeon and Bokes(2012)]{gedeon2012delayed}
Tom{\'a}{\v{s}} Gedeon and Pavol Bokes.
\newblock Delayed protein synthesis reduces the correlation between mrna and
  protein fluctuations.
\newblock \emph{Biophysical journal}, 103\penalty0 (3):\penalty0 377--385,
  2012.

\bibitem[Alvarez et~al.(2009)Alvarez, Luengo, and Lawrence]{alvarez2009latent}
Mauricio Alvarez, David Luengo, and Neil~D Lawrence.
\newblock Latent force models.
\newblock In \emph{Artificial Intelligence and Statistics}, pages 9--16, 2009.

\bibitem[Lawrence et~al.(2007)Lawrence, Sanguinetti, and
  Rattray]{lawrence2007modelling}
Neil~D Lawrence, Guido Sanguinetti, and Magnus Rattray.
\newblock Modelling transcriptional regulation using gaussian processes.
\newblock In \emph{Advances in Neural Information Processing Systems}, pages
  785--792, 2007.

\bibitem[Titsias et~al.(2012)Titsias, Honkela, Lawrence, and
  Rattray]{titsias2012identifying}
Michalis~K Titsias, Antti Honkela, Neil~D Lawrence, and Magnus Rattray.
\newblock Identifying targets of multiple co-regulating transcription factors
  from expression time-series by bayesian model comparison.
\newblock \emph{BMC systems biology}, 6\penalty0 (1):\penalty0 53, 2012.

\bibitem[L{\'o}pez-Lopera et~al.(2019)L{\'o}pez-Lopera, Durrande, and
  Alvarez]{lopez2019physically}
Andr{\'e}s~Felipe L{\'o}pez-Lopera, Nicolas Durrande, and Mauricio~Alexander
  Alvarez.
\newblock Physically-inspired gaussian process models for post-transcriptional
  regulation in drosophila.
\newblock \emph{IEEE/ACM transactions on computational biology and
  bioinformatics}, 2019.

\bibitem[Barenco et~al.(2006)Barenco, Tomescu, Brewer, Callard, Stark, and
  Hubank]{barenco2006ranked}
Martino Barenco, Daniela Tomescu, Daniel Brewer, Robin Callard, Jaroslav Stark,
  and Michael Hubank.
\newblock Ranked prediction of p53 targets using hidden variable dynamic
  modeling.
\newblock \emph{Genome biology}, 7\penalty0 (3):\penalty0 R25, 2006.

\bibitem[Giampieri et~al.(2011)Giampieri, Remondini, De~Oliveira, Castellani,
  and Li{\'o}]{giampieri2011stochastic}
E~Giampieri, D~Remondini, L~De~Oliveira, G~Castellani, and P~Li{\'o}.
\newblock Stochastic analysis of a mirna--protein toggle switch.
\newblock \emph{Molecular bioSystems}, 7\penalty0 (10):\penalty0 2796--2803,
  2011.

\bibitem[Santill{\'a}n(2008)]{santillan2008use}
Moises Santill{\'a}n.
\newblock On the use of the hill functions in mathematical models of gene
  regulatory networks.
\newblock \emph{Mathematical Modelling of Natural Phenomena}, 3\penalty0
  (2):\penalty0 85--97, 2008.

\bibitem[Hoffman and Gelman(2014)]{hoffman2014no}
Matthew~D Hoffman and Andrew Gelman.
\newblock The no-u-turn sampler: adaptively setting path lengths in hamiltonian
  monte carlo.
\newblock \emph{Journal of Machine Learning Research}, 15\penalty0
  (1):\penalty0 1593--1623, 2014.

\bibitem[Hafner et~al.(2017)Hafner, Stewart-Ornstein, Purvis, Forrester, Bulyk,
  and Lahav]{hafner2017p53}
Antonina Hafner, Jacob Stewart-Ornstein, Jeremy~E Purvis, William~C Forrester,
  Martha~L Bulyk, and Galit Lahav.
\newblock p53 pulses lead to distinct patterns of gene expression albeit
  similar dna-binding dynamics.
\newblock \emph{Nature structural \& molecular biology}, 24\penalty0
  (10):\penalty0 840, 2017.

\end{thebibliography}
\bibliographystyle{unsrtnat}

\end{document}